\documentclass[prl,preprint,showpacs,preprintnumbers,amsmath,amssymb]{revtex4}

\usepackage{graphicx}
\usepackage{dcolumn}
\usepackage{bm}
\usepackage{epstopdf}
\begin{document}

\preprint{APS/123-QED}
\title{Scaling exponents and probability distributions of DNA end-to-end distance}

\author{Francesco Valle}
\author{M\'{e}lanie Favre}
\author{Paolo De Los Rios*}
\author{Angelo Rosa$\dagger$}
\author{Giovanni Dietler}
\affiliation{Laboratoire de Physique de la Mati\`{e}re Vivante, IPMC,\\
*Laboratoire de Biophysique Statistique, ITP,\\
$\dagger$Institut de Math\'{e}matique B,\\
{\'E}cole Polytechnique F\'{e}d\'{e}rale de Lausanne (EPFL), CH-1015 Lausanne, Switzerland}
\date{\today}
\begin{abstract}
Correlation length exponent $\nu$ for long linear DNA molecules was determined by
direct measurement of the average end-to-end distance as a
function of the contour length $s$ by means of atomic force
microscopy (AFM). Linear DNA, up to 48'502 base pairs (bp), was
irreversibly deposited from a solution onto silanized mica and
imaged in air.  Under the adsorption conditions
used, the DNA is trapped onto the surface without any
two-dimensional equilibration. The measured exponent is $\nu = 0.589 \pm 0.006$, in agreement with the theoretical 3D value of $\nu = 0.5880 \pm 0.0010$. The persistence length $\ell_p$ of DNA was estimated to be 44$\pm$3 nm, in agreement with the literature values. The distribution of the
end-to-end distances for a given contour length $s$ and the exponents
characterizing the distribution were determined for different $s$. For $s$ smaller or comparable to $\ell_p$, a delta function like distribution was observed, while for larger $s$, a probability distribution of the type $x^{d-1}x^g e^{-bx^\delta}$ was observed with $g=0.33\pm0.22$ and $\delta=2.58\pm0.76$. These values are compared to the theoretical exponents for Self-Avoiding Walk (SAW): namely $g=\frac{\gamma-1}{\nu}$ and $\delta=(1-\nu)^{-1}$. So for $d=2$, $g\approx0.44$ and $\delta=4$, while for $d=3$, $g\approx0.33$ and $\delta\approx2.5$. The derived entropic exponent $\gamma$ is $\gamma=1.194\pm0.129$. The present data indicate that the DNA behaves on large length scales like a 3 dimensional SAW.
\end{abstract}


\maketitle

The statical properties of polymers are the focus of a strong research effort. In the limit of very long and perfectly flexible linear polymers, the situation is rather clear and the principles were laid down some time ago \cite{degennes1}. Self-Avoiding Walks (SAWs) describe the properties of very long polymers in 2 and 3 dimensions and the main results can be summarized in the scaling properties of polymers which were theoretically as well as experimentally confirmed \cite{degennes1}. Concerning DNA, the situation is more complex due to the elastic properties of the double helix, its polyelectrolytic properties and its persistence length $\ell_p$(see \cite{ghosh,winkler2}). Experimentally, the dynamics and statics of a purely two dimensional linear DNA chain was investigated by Maier et al. \cite{radler1} finding results in agreement with the theoretical predictions.  Moreover, Rivetti et al. \cite{rivetti1} used Atomic Force Microscopy (AFM) to investigate statistical properties of DNA yielding information about the persistence length, the kinetics and the mode of adsorption on a substrate.  Local changes in rigidity, curvature and/or topology of a DNA molecule induced by chemical compounds or by DNA binding proteins \cite{lyubchenko1,rivetti2,jovin1,Viglasky, samori1,Viglasky1} have been also studied. However, the three dimensional conformation of a flexible biopolymer is not directly delivered by the AFM, but we show here that such information can be extracted from the AFM images.

Here we present the determination of the end-to-end distance and its probability distribution as a function of the contour length $s$ of DNA on a range spanning 4 orders of magnitude: from $s=1\ nm$ to $s=10'000\ nm$, a range which spans values both smaller and much larger than the persistence length $l_p$. This is only possible since the atomic force microscope \cite{binnig} has a very high spatial resolution but it can also image biomolecules on very larger scales. Information at the single molecule level has become available on the basis of the images of DNA molecules adsorbed onto a
surface.

In the present work, the irreversible adsorption of DNA on a flat
surface is investigated by measuring the mean end-to-end distance
as a function of the length of the polymer contour. The
experimental results show that there are two scaling regimes. At
short length scales (i.e. smaller than the persistence length
$\ell_p$), the DNA behaves like a rigid rod. On length scales bigger
than $\ell_{p}$, a 3 D behavior is observed with a Flory exponent
$\nu$=0.589$\pm$0.006. Furthermore, working with images of single
DNA molecules allowed us to determine the distribution of the
end-to-end distances for a given contour length, an
information usually unavailable. The form of the distribution depends on the flexibility of the polymer and on the contour length considered. Although several theoretical estimations of the distributions
exist \cite{frey,thirumalai,winkler}, so far we are not aware of any direct experimental
measurements of these parameters.

Linear DNA was prepared from a solution of $\lambda$-phage DNA,
48'502 base pairs (bp) long, cleaved by restriction enzymes to
give a mixture of lengths from 1'503 bp up to the maximum 48'502
bp. DNA molecules were prepared in a buffer solution of 10 mM
Tris-HCl, pH 7.6, 1 mM EDTA with DNA concentrations ranging from
0.5 $\mu$g/ml to 2 $\mu$g/ml. The Debye screening length $1/k$ is $2\ nm$ \cite{volo}. The
substrates (freshly cleaved mica) were positively charged by
exposing them to 3-aminopropyltriethoxy silane (APTES) vapors
during 2 hours at room temperature in a dry atmosphere
\cite{lyubchenko2}. A 10 $\mu$l drop of a DNA solution was
deposited onto the substrate surface during 10 minutes and then
rinsed with ultra-pure water. The sample was finally blown
dry with clean air. The DNA images were recorded by means of an AFM operated in tapping mode, in order
to reduce the effect of lateral forces during scanning of the
surface \cite{Garcia}. We checked that the sample remains stable
for weeks if kept in dry atmosphere, proving the irreversibility
of the adsorption. In figure 1 are depicted four DNA images. From
such pictures, the contour of about 60 DNA molecules was
digitized using a specially designed software\cite{Marek}, which allows to track the
molecule backbone and to extract the coordinates of the polymer
contour. The first digitization, which may lead to noisy and not
equidistant coordinates, was subsequently smoothed using the Snake
algorithm \cite{snake1}. Several tests were performed with
molecules of known length (DNA plasmids) to check the procedure\cite{Marek}.
\begin{figure}[!]
\includegraphics[width=8cm]{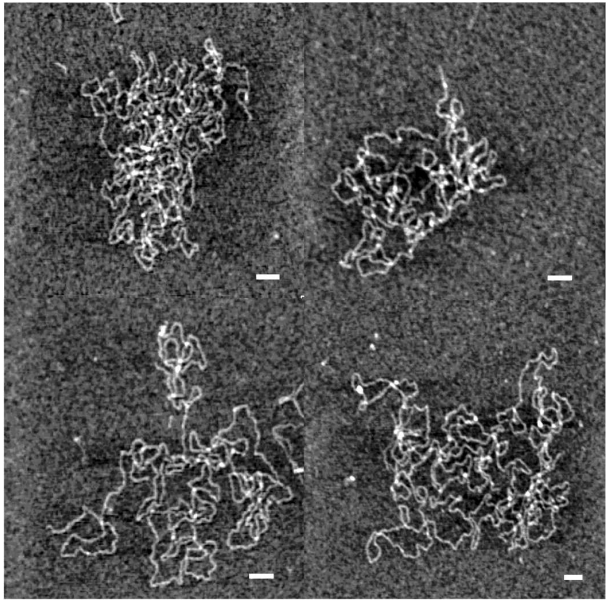}
\caption{\label{fig_1} a) Tapping mode images of linear DNA
molecule from an enzymatic digested lambda DNA. The original DNA
has 48'502 base pairs, here we show some shorter fragments. The
scale bar represents 100 nm}
\end{figure}
The analysis of the end-to-end distance $R(s)$ for the DNA
molecules as a function of the contour length $s$ was done by
moving a window of length $s$ along the contour of the DNA
molecule for values of $s$ going from the minimal segment value
$s_{min}$ = 2 nm to the total DNA length. The small values of $s$
have a better statistical error, since they occur more often than
the maximum value. The $R(s)$ values  were then averaged
over all the molecules to yield the mean end-to-end
$<R(s)>$, which is plotted in Figure \ref{fig_2}. One expects that $R(s)$ scales with a single power
law:
\begin{equation}
<R(s)> \sim s^{\nu}\nonumber
\end{equation}
It is clear from Figure \ref{fig_2}  that the above power law is
not respected over the whole interval of $s$. A two power law
function has to be used in order to fit the data:
\begin{equation}
<R(s)> \sim
\left(\frac{s}{\ell}\right)^{\nu_o}\left(1+\frac{s}{\ell}\right)^{\nu_1-\nu_o}\label{fit}
\end{equation}
where $s$ is the contour length, $\ell$ a cross-over length
between the two power laws (it will be shown to correspond to the
persistence length $\ell_p$), and $\nu_o$ and $\nu_1$ the scaling (critical) exponents.
The fit to the data gives the following results:
\begin{eqnarray}
\nu_o &  = & 1.030 \pm 0.017 \nonumber\\
\nu_1&  = & 0.589 \pm 0.006 \nonumber\\
\ell & = & (44 \pm 3)\ nm\nonumber
\end{eqnarray}
\begin{figure}[!]
\includegraphics[width=8cm]{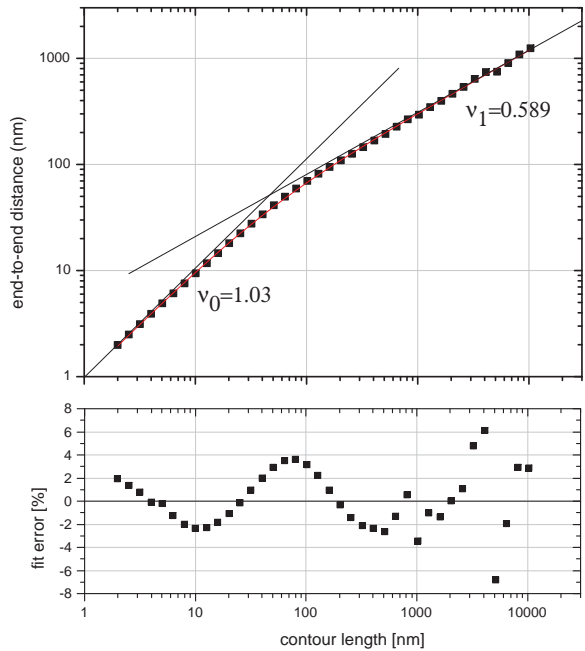}
\caption{\label{fig_2} Representation in double logarithmic scale
of the end-to-end distance vs the contour length. Two distinct
power laws are evident; the data have been fitted using equation
\ref{fit}.}
\end{figure}
These results can be interpreted as follows. The value of $\ell$
is in good agreement with previously reported measurements of the persistence length of DNA
performed by microscopy techniques \cite{stasiak1,henderson1}. We can therefore identify $\ell$ with $\ell_p$, which in turn leads to a simple and intuitive interpretation of the two scaling regimes. For  $s < \ell_p$ DNA behaves as a rigid rod and the end-to-end distance scales linearly with the contour length of the polymer ($\nu_o=1.030$). For $s>\ell_p$ the scaling exponent ($\nu_1=0.589$) agrees with the best numerical estimations of the exponents by renormalization group
techniques ($0.5880\pm0.0010$ \cite{zinnjustin}) and with
experimental values for synthetic polymers by scattering methods\cite{degennes2,japan}.
Thus, the adsorbed DNA behaves as a three-dimensional
polymer, which does not undergo any two-dimensional equilibration upon absorption. This is also evident from the images of Figure \ref{fig_1}, where
 the DNA molecules show a large number of crossings. The euclidean dimension of the surface ($d=2$) onto which the molecule is projected, is larger than the fractal dimension
 $(d_f=1/\nu_1\approx1.7)$ of the DNA, explaining the
conservation of the 3D exponent upon adsorption on a
surface \cite{mandelbrot}. We conclude that the $\lambda$-DNA images represent some form of a
two-dimensional projection of their three dimensional bulk conformation.

\begin{figure}[!]
\includegraphics[width=8cm]{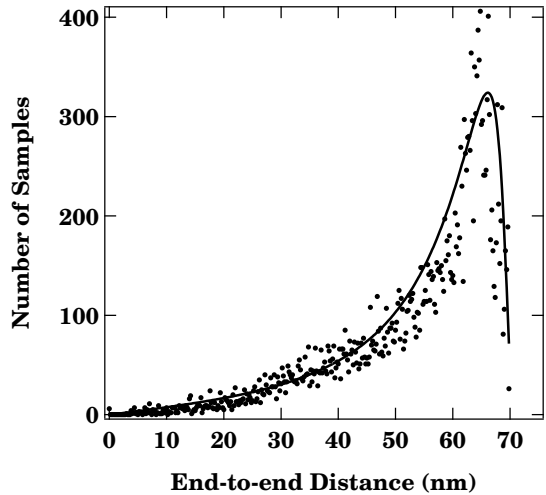}
\caption{\label{fig_3} Histogram representing the distribution of
the end-to-end distance for a contour length $s_0=75\ nm$. The continous line is a fit to equation equation \ref{winkler2D}.}
\end{figure}

Furthermore, we can measure the probability distribution of the end-to-end distance for a determined contour length $s$ as it can be extracted from the microscopic conformation of
each molecule. Two regimes are possible in respect to the persistence length: $s\lesssim\ell_p$, and $s>\ell_p$. For $s\lesssim\ell_p$, one expects an almost delta function distribution, since the polymer chains do not bend over
these length scales. For $s\approx\ell_p$, the distribution is still narrow but not as peaked as for the previous case, since the molecules starts to enter the semiflexible regime: an example is given in Figure \ref{fig_3} for a contour length of $s_o=75\ nm$ ($\approx 2 \ell_p$). This distribution was fitted with Winkler's equation \cite{winkler} 

\begin{equation}
f(s)=a \frac{s\ e^{-\frac{s_o}{8 \ell_p (1-(s/s_o)^2)}}}{\left[1-(s/s_o)^2\right]\left[2-(s/s_o)^2\right]^2}\label{winkler2D}
\end{equation}

\begin{figure}[!]
\includegraphics[width=8cm]{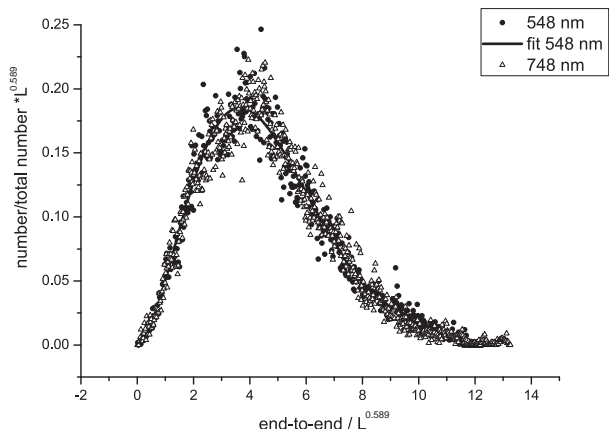}
\caption{\label{fig_4} Histogram representing the distribution of
the end-to-end distance for two different contour length (548 nm
filled circles, 748 nm open triangles) and how the collapse onto
each other once they have been rescaled using the scaling exponent $\nu_1=0.589$ measured in the present experiments. The solid line represents the
fit of the 548 nm distribution using equation \ref{distribution}.}
\end{figure}

The fits shown in Figure \ref{fig_3} gives $\ell_p=46.6\ nm$ and $s_o=71.4\ nm$, in good agreement with the literature value for the persistence length and with the nominal total length of $75\ nm$. For contour lengths $s_o>>\ell_p$, the distribution changes dramatically and was determined for  34 different contour lengths, ranging from $200\ nm$ to $4'600\ nm$. For longer contour lengths, the distributions are difficult to determine because of the reduced number of samples available and only the average end-to-end distance can be given like in Figure \ref{fig_1}. In Figure \ref{fig_4} we show the distributions for $s_0=548\ nm$ ($\approx12\ell_p$) and $s_o=748\ nm$($\approx17\ell_p$). The histograms of figure \ref{fig_4} have been rescaled with
$s^{\nu_{1}}$ and the way they
collapse onto each other is a good "\textit{a posteriori}"
confirmation of the power law previously determined (Figure
\ref{fig_2}). Starting from a SAW model \cite{degennes1,grosberg}, the distribution probability of the end-to-end distance as a function of the contour length $s$ becomes:

\begin{equation}
f(s)= as^{d-1}s^{\sigma}e^{-bs^{\delta}}\label{distribution}
\end{equation}
The two exponents characterizing the distributions were determined from fits of the histograms for the 34 different contour lengths between $200\ nm$ and $4'600\ nm$;
their averages are:
\begin{eqnarray}
\sigma = 0.33 \pm 0.22\nonumber\\
\delta = 2.58 \pm 0.76\nonumber
\end{eqnarray}

These values have to be compared to those characterizing
the corresponding  two or three dimensional distributions of the end-to-end distances
for  a SAW (for $d=2$, $\sigma=0.44$, $\delta=4$ and for $d=3$, $\sigma=0.33$, $\delta=2.43$). The present results indicate, that the DNA chains behave like a 3 dimensional SAW.


First, it is a confirmation that in this range of DNA lengths
($s_o\sim5-100$ persistence lengths) the end-to-end distribution is matching a pure SAW distribution. Under the conditions used in our preparation, the DNA adsorption is strong and DNA is quenched on the surface. No equilibration is taking place in 2 dimensions. The problem of "trapping" or "equilibration" of DNA onto different surfaces has been already studied by Rivetti et al. \cite{rivetti1}, using short DNA fragments which were then analyzed with the Worm Like Chain (WLC) model. In their work they
saw large deviation from the expected three-dimensional WLC
already for fragments of 6 kbp. In our work the 3D power law fits
the experimental data up to $\approx$30 kbp or $230\ \ell_p$, with
no \emph{a priori} assumption about the dimensionality of the
final conformation. Compared to the optical images of fluorescently marked polymers
\cite{radler1,radler2}, the high resolution of AFM allows to
perform the analysis on segments as short as few nanometers up to
$10'000\ nm$ over 4 decades of lengths allowing to observe the
transition from stiff to SAW polymer behavior.

Further experiments should be performed by varying the salt concentration in order to determine the contribution of the electrostatic persistence length to the total persistence length (see \cite{winkler2}). However, the use of high salt concentrations will also change the deposition process \cite{rivetti1} and the adsorption might not be irreversible. A certain degree of 2 dimensional equilibration might take place, influencing the persistence length \cite{revet1, rivetti1}. Moreover, the theory of polyelectrolytes with stiffness should be used if salt and other parameters could be varied\cite{ghosh,winkler2}. In these theories a more complex behavior is predicted: a crossover from stiff rod behavior to SAW through a region of Gaussian behavior is considered. At present our data do not allow to make such a detailed test.
\acknowledgements{We would like to thank A. Stasiak, J. Prost, L. Peliti, R. Metzler, and R. Winkler for suggestions, comments and important discussions on the interpretation of the data. G.D. thanks the Swiss National Science Foundation for support through the grant Nr.  2100-063746.00.}

\end{document}